\documentclass[11pt,a4paper]{article}
\pdfoutput=1 

\usepackage[table]{xcolor}
\usepackage{colortbl}
\RequirePackage{ifpdf} 
\usepackage{amsmath} 
\usepackage{mathtools}
\usepackage{cases}

\usepackage{jheppub}
\usepackage{pstricks}
\usepackage[final]{pdfpages} 
\usepackage{ifpdf} 
\usepackage{slashed}

\usepackage{color}
\usepackage{xcolor}
\definecolor{urlblue}{rgb}{0.2,0.4,0.7}
\definecolor{citegreen}{rgb}{0,0.6,0.2}
\definecolor{linkred}{rgb}{0.9,0.2,0.1}
\usepackage{hyperref}

\usepackage{graphics}
\usepackage{etoolbox} 
\usepackage{fixmath}
\usepackage{psfrag}

\usepackage{notoccite} 

\usepackage{amsfonts}
\usepackage{autobreak}
\usepackage{marginnote}

\def\beq{\begin{equation}}
\def\eeq{\end{equation}}
\def\bea{\begin{eqnarray}}
\def\eea{\end{eqnarray}}

\def\qgraf{{\fontfamily{qcr}\selectfont
QGRAF}}
\def\form{{\fontfamily{qcr}\selectfont
FORM}}
\def\litered{{\fontfamily{qcr}\selectfont
LiteRed}}
\def\fire{{\fontfamily{qcr}\selectfont
FIRE5}}
\def\cpp{{\fontfamily{qcr}\selectfont
C++}}
\def\mint{{\fontfamily{qcr}\selectfont
Mint}}
\def\arXiv{{\fontfamily{qcr}\selectfont
arXiv}}

\title{Four-particle scattering amplitudes in QCD at NNLO to higher orders in the dimensional regulator}

\author{
Taushif Ahmed$^a$,
Johannes Henn$^a$ and
Bernhard Mistlberger$^b$
}

\affiliation{
$^a$ Max-Planck-Institut f{\"u}r Physik, Werner-Heisenberg-Institut, 80805 M{\"u}nchen, Germany\\
$^b$ Center for Theoretical Physics, Massachusetts Institute of Technology, Cambridge, MA 02139, USA
}

\emailAdd{taushif@mpp.mpg.de}
\emailAdd{henn@mpp.mpg.de}
\emailAdd{bernhard.mistlberger@gmail.com}

\preprint{MIT-CTP/5130, MPP-2019-129}

\abstract{
We compute all helicity amplitudes for four particle scattering in massless QCD with $n_f$ fermion flavours to next-to-next-to-leading order (NNLO) in perturbation theory. 
In particular, we consider all possible configurations of external quarks and gluons. 
We evaluate the amplitudes in terms of a Laurent series in the dimensional regulator 
to the order required for future next-to-next-to-next-to-leading order (N$^3$LO) calculations.
The coefficients of the Laurent series are given in terms of harmonic polylogarithms that can readily be evaluated numerically. 
We present our findings in the conventional dimensional regularisation and in the t'Hooft-Veltman schemes.
}

\begin{document}
\allowdisplaybreaks[4]
\unitlength1cm
\keywords{QCD, Four-point, Two-loop, Helicity, Scattering amplitudes}
\maketitle

\section{Introduction}
\label{sec:intro}

The Large Hadron Collider (LHC) provides us with the opportunity to test our understanding of fundamental physics at unprecedented energies.
Of particular importance are processes that shed light on the structure of strong interactions.
Scattering processes involving the strong interaction often result in the production of sprays of collimated hadrons - so-called jets. 
One of the most prominent examples of such an observable is the production cross section for two jets at the LHC.
This process can be measured to astounding precision~\cite{Aaboud:2017wsi,Khachatryan:2016wdh,Sirunyan:2017skj}.
To draw conclusions about the fundamental interactions we require equally precise predictions for such scattering observables. 

The observables associated with the production of two jets were computed at next-to-leading order (NLO)~\cite{Gao:2012he,Alioli:2010xa,Giele:1994gf,Ellis:1992en} and even at NNLO~\cite{Currie:2017eqf,Gehrmann-DeRidder:2019ibf,Currie:2018oxh,Czakon:2019tmo} in perturbative QCD. 
The importance of this process and the wealth of data collected by the LHC is inspiring to go beyond the existing accuracy and perform a calculation at N$^3$LO in QCD perturbation theory.
The key ingredients for this formidable goal are virtual four point amplitudes of massless, on-shell QCD partons. 
In particular, working in dimensional regularisation in the modified minimal subtraction scheme it is necessary to obtain tree level, 1-loop, 2-loop and 3-loop amplitudes to $6^{th}$, $4^{th}$, $2^{nd}$ and $0^{th}$ order in the dimensional regulator, respectively.
It is the goal of this article to provide the foundation of such a computation by obtaining all required amplitudes to the desired power in the dimensional regulator through NNLO. 
Previous results for these scattering amplitudes were obtained to lower power in the dimensional regulator in refs.~\cite{Anastasiou:2000kg,Anastasiou:2000ue,Bern:2000dn,Glover:2001af,Anastasiou:2001sv,Bern:2002tk,Bern:2003ck,DeFreitas:2004kmi,DeFreitas:2004aj,Abreu:2017xsl,Glover:2003cm,Glover:2004si,Broggio:2014hoa}. Very recently, the four-gluon amplitude at three-loop level in the planar limit in Pure-Yang Mills theory is computed in ref.~\cite{Luo2019}.

 In this article, we compute all relevant four-point amplitudes for the scattering of on-shell, massless QCD partons up to two loops.
 In particular we consider the scattering processes
\begin{align}
    &gg \rightarrow gg\,,\nonumber\\
    &q{\bar q} \rightarrow gg\,,\nonumber\\
    &q{\bar q} \rightarrow q{\bar q}\,,\nonumber\\
    &q{\bar q} \rightarrow Q{\bar Q}.
\end{align}
Here, $g$ are gluons, and $q,Q$ represent quarks of different flavours. 
All further scattering processes can be obtained by crossing parton momenta.
We explain this procedure in detail.

In our computation we apply dimensional regularisation in the $\overline{\text{MS}}$ scheme.
This means extending the space time dimension to $d=4-2\epsilon$ to regulate infrared and ultraviolet divergences.
There are several variants of this regularisation scheme mainly differing by the number of assigned polarisation to massless gauge bosons.
In conventional dimensional regularisation (CDR), one assigns $2-2\epsilon$ polarisation states to all internal as well as external gluons. 
In the helicity approach, external gluons are associated with 2 physical polarisation states and some freedom exists regarding internal, virtual gluons.
In 't Hooft Veltman (HV) scheme~\cite{tHooft:1972tcz}, virtual gluons are assigned $2-2\epsilon$ polarisations whereas in four dimensional helicity (FDH)~\cite{Bern:1991aq} they are assigned exactly two polarisations degree of freedom. 
We derive our amplitudes using projection operators and our results are valid for arbitrary space time dimension as in CDR. 
Furthermore, we relate our results to HV scheme amplitudes and present them making use of the spinor-helicity formalism~\cite{Berends:1981rb,DeCausmaecker:1981wzb,Xu:1986xb}.
In order to convert amplitudes in the HV scheme to the FDH scheme see for example ref.~\cite{Broggio:2015dga}.


The structure of ultraviolet, infrared and collinear singularities of massless QCD scattering amplitudes is well understood~\cite{Catani:1998bh,Sterman:2002qn,Aybat:2006wq,Aybat:2006mz,Becher:2009cu,Gardi:2009qi,Almelid:2015jia,Almelid:2017qju}.
We briefly review this structure and verify that our amplitudes adhere to these universal principles. 
As a consequence of our computation, it is now possible to predict all poles in the dimensional regulator appearing in the equivalent three loop, four point amplitudes for massless parton scattering.

The main result of this article are analytic formulae for all four parton scattering amplitudes in massless QCD to second order in the strong coupling.
We present this results in electronic form together with the \arXiv~submission of this article. 
The basic building blocks of the amplitudes are well known functions - so-called harmonic polylogarithms (HPLs)~\cite{Remiddi:1999ew} that are described briefly below.

The paper is organised as follows. 
In section~\ref{sec:setup}, we discuss the basic ingredients of our computation.
In section~\ref{sec:four-gluon}, the four-gluon amplitude is described. 
The quark-gluon scattering amplitude is discussed in the next section~\ref{sec:quark-gluon}. 
In section~\ref{sec:quark}, the four-quark amplitudes with identical as well as different flavours are discussed. 
The ultraviolet renormalisation, along with the infrared factorisation, is discussed in section~\ref{sec:IR}. 
The consistency checks which are performed to ensure the correctness of the results are tabulated in section~\ref{sec:checks}. 
We conclude in section~\ref{sec:concl}. 

\section{Setup}
\label{sec:setup}
In this section, we discuss the basic ingredients for our computation of scattering amplitudes of four massless particles in massless QCD.
We label the momenta of the four scattering particles by $p_1,\dots,p_4$ and regard all of them as incoming.
Let us define the following abbreviations
\bea
s_{ij}&=&(p_i+p_j)^2\hspace{1cm}{\rm for}\hspace{1cm}i\neq j\,,\nonumber\\
p_{i_1\dots i_n}&=&p_{i_1}+\dots+p_{i_n}.
\eea
At four points with only massless external legs momentum conservation dictates that
\bea
&p_1+p_2+p_3+p_4=0\,,&\nonumber\\
&s_{12}+s_{13}+s_{23}=0.&
\eea
Exploiting momentum conservation we can express all our amplitudes by one dimensionless ratio $x$ and absorb all the dependence on the energy dimension into the variable $s$.
\bea
s&=&s_{12}\,,\hspace{1cm} x=-\frac{s_{13}}{s_{12}}.
\eea
For convenience we set $s=1$ as it can be easily reconstructed by dimensional analysis.
Furthermore, for the purpose of this article we assume that $x\in [0,1]$ which corresponds to a region of physical $2\rightarrow 2$ scattering.

We perform a perturbative expansion of our amplitudes in the strong coupling constant.
The bare strong coupling constant is related to the renormalised strong coupling constant by
\bea
\label{eq:coup}
\frac{\alpha_0}{\pi}&=&\frac{g^2}{16\pi^2} e^{\epsilon \gamma_E} \left( \frac{\mu^2}{4\pi}\right)^{-\epsilon}Z, \nonumber\\
a_R& =&\frac{g^2}{16\pi^2}. 
\eea 
Here, $\epsilon$ is the dimensional regulator introduced through $d=4-2\epsilon$ with the space-time dimension $d$, and $a_R$ is the renormalised strong coupling constant. 
The Euler-Mascheroni constant is denoted by $\gamma_E$ and the renormalisation scale is denoted by $\mu$. 
For most of the article we work in conventional dimensional regularisation (CDR) assigning $2-2\epsilon$ polarisations to each gluon (internal and external).
In our computation we assume $n_f$ massless quark flavours.
We furthermore choose $\mu^2=s$ as a convenient scale to represent our amplitudes. 
The renormalisation factor $Z$ is discussed in section~\ref{sec:IR}. 
We furthermore introduce the expansion paramter $a=a_R Z$.   
With this our amplitudes can be written as
\beq
\label{eq:amp}
\mathcal{A}_{X}(s,x,\epsilon)= \sum\limits_{i=0}^\infty  a^i \left(\frac{s}{\mu^2}\right)^{-i \epsilon}\mathcal{A}^{(i)}_{X}(x,\epsilon).
\eeq  
Here, $X$ represents the particles that are scattering.
For example, $X=ggq\bar q$ indicates an amplitude where two gluons with momenta $p_1$ and $p_2$ are scattering with a quark  and anti-quark with momenta $p_3$ and $p_4$, respectively.

In order to compute all required amplitudes we start by generating the necessary Feynman diagrams using \qgraf~\cite{Nogueira:1991ex}. 
We then perform colour and spinor algebra using private \cpp~ code. 
We project all Lorentz, spinor and colour structures to scalar amplitudes using projectors, see for example refs.~\cite{Chen:2019wyb,Gehrmann:2013vga,Peraro:2019cjj}.
Subsequently, we reduce the  resulting loop integrands to a basis of so-called master integrals by means of the Laporta algorithm~\cite{Laporta:2001dd} implemented in a private code. 
In a complementary implementation, a set of in-house routines written in symbolic manipulating program \form~\cite{Vermaseren:2000nd,Ruijl:2017dtg} is utilised to perform the spinor and colour algebra. 
Then all the scalar loop integrals are reduced to a set of master integrals using integration-by-parts~\cite{Tkachov:1981wb,Chetyrkin:1981qh} identities with the help of publicly available programs \litered~\cite{Lee:2008tj,Lee:2012cn} along with \mint~\cite{Lee:2013hzt} and \fire~\cite{Smirnov:2014hma}.
We then compute all required master integrals as a Laurent expansion in the dimensional regulator using the method of differential equations~\cite{Kotikov:1991pm,Bern:1993kr,Gehrmann:1999as,Henn:2013pwa}.
In particular we use the same basis of integrals as was used for the computation of refs.~\cite{Henn:2016jdu,Henn:2019rgj}.
Inserting the master integrals into our reduced amplitudes we obtain our final results.
We represent our amplitudes in a particular basis of colour structures and Lorentz or spinor structures that we detail below.

We will express our result for amplitudes at different orders in the coupling expanded in the dimensional regulator in terms of harmonic poly-logarithms (HPLs)~\cite{Remiddi:1999ew}.
HPLs are defined by
\beq
H(b_n,\dots,b_1;x)=(-1)^{\frac{1}{2}(|b_n|+b_n)}\int_0^x \frac{dx^\prime}{x^\prime-b_n}H(b_{n-1},\dots,b_1;x^\prime),\hspace{1cm}b_i\in\{0,-1,1\}.
\eeq
We refer to $x$ as the argument of a HPL and to the $b_i$ as its indices. 
If the right-most indices are zero, the HPL would be divergent and we work with the regulated definition
\beq
\label{eq:reglogdef}
H(0,\dots,0;x)=\frac{1}{n!}\log^n(x).
\eeq
The HPLs are widely used in the literature and their properties are well understood, see for example refs.~\cite{Goncharov:1998kja,Panzer:2014gra,Duhr:2011zq,Panzer:2014caa,Duhr:2012fh,Maitre:2005uu}.

\section{The Four-Gluon Amplitude}
\label{sec:four-gluon}
\noindent
\begin{figure}[h!]
\begin{center}
\includegraphics[width=0.30\textwidth]{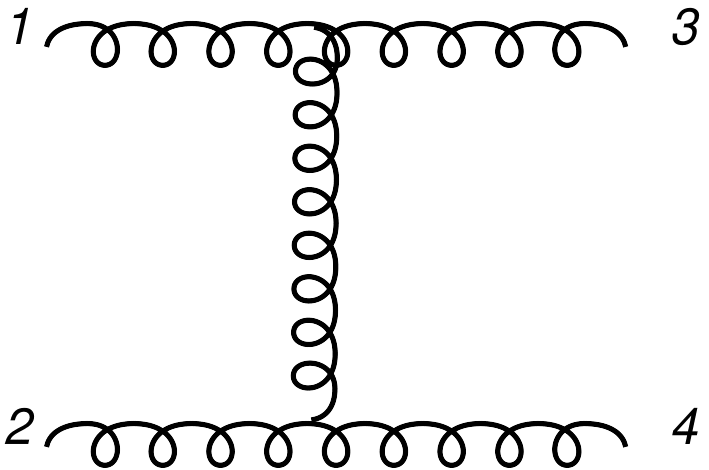}
\end{center}
\caption{Example of a leading order Feynman diagram for the scattering process of four gluons.}
\end{figure}
The amplitude for the scattering of four gluons can be decomposed into different colour and Lorentz tensors. 
We begin by introducing our basis of tensor structures.

In order to define colour tensors with adjoint colour indices we introduce traces of fundamental generators $T^a_{ij}$.
We abbreviate
\beq
tr(T^{a_1}T^{a_2}T^{a_3}T^{a_4})=tr(1234).
\eeq
The vector of colour structures that we will use to express the amplitude is given by 
\beq
\vec{c}^{~a_1a_2a_3a_4}_{gggg}=\left(\begin{array}{c} 
tr(1234)+tr(1432)\\
tr(1243)+tr(1342)\\
tr(1423)+tr(1324)\\
tr(12)tr(34)\\
tr(13)tr(24)\\
tr(14)tr(23)\\
\end{array}\right).
\eeq


The Feynman integrand for our amplitude is a Lorentz tensor contracted with polarisation vectors $\epsilon^\mu_i(p_i)$. 
In order to represent the Lorentz tensor we make an ansatz in terms of all $p_i^{\mu_i}$ and $g^{\mu_i \mu_j}$. 
Since we are considering an amplitude with four external gluons, our ansatz has tensor rank four. 
This yields in full generality 138 independent coefficients.
To restrict ourselves to the physical subspace of the amplitude we choose a light-like axial gauge (see for example ref.~\cite{Gehrmann:2013vga}) with the gauge conditions
\bea
\epsilon^\mu_i p_{i,\mu}=\epsilon^\mu_i p_{i-1,\mu}=0.
\eea
We identify $p_0^\mu=p_4^\mu$.
In this gauge, the polarisation sum that has to be carried out to obtain squared matrix elements takes the form
\beq
\sum_{\text{polarisations}} \epsilon^\mu_i \epsilon^{*\,\nu}_i =g^{\mu\nu} -\frac{2}{s_{i,i-1}}\left(p_i^\mu p_{i-1}^\nu-p_i^\nu p_{i-1}^\mu\right).
\eeq
Effectively, this polarisation sum annihilates every vector $p_i^\mu$ or $p_{i-1}^\mu$ in the amplitude. Only polarisations transverse to the two chosen vectors survive. 
We can anticipate the projection that would be performed when interfering the amplitude with its complex conjugate to produce squared matrix elements.
We do so by replacing 
\beq
\epsilon_i^\mu\rightarrow \left[ g^{\mu\nu} -\frac{2}{s_{i,i-1}}\left(p_i^\mu p_{i-1}^\nu-p_i^\nu p_{i-1}^\mu\right) \right]\epsilon_i^\nu
\eeq
in our ansatz for the general amplitude. 
This procedure effectively annihilates all but 20 tensor structures in our ansatz. 
After simple orthogonality checks we find that only 10 of the remaining structures are linearly independent. 
In order to represent those ten physical structures we define
\beq
P_{ij}^\mu=\frac{p_j^{\mu }}{s_{j,i}}-\frac{p_i^{\mu } s_{j,i-1}}{s_{i,i-1} s_{j,i}}-\frac{p_{i-1}^{\mu }}{s_{i,i-1}}
\eeq
and 
\bea
G_{ ij}^{\mu\nu}&=&g^{\mu  \nu }-\frac{2 p_{i-1}^{\mu } p_j^{\nu } s_{j-1,i}}{s_{i,i-1} s_{j,j-1}}+\frac{2 p_{i-1}^{\mu } p_{j-1}^{\nu } s_{j,i}}{s_{i,i-1} s_{j,j-1}}+\frac{2
   p_i^{\mu } p_j^{\nu } s_{j-1,i-1}}{s_{i,i-1} s_{j,j-1}}\nonumber\\
   &+&\frac{2 p_i^{\mu } p_{j-1}^{\nu } s_{j,i-1}}{s_{i,i-1} s_{j,j-1}} -\frac{2 p_{i-1}^{\mu }  p_i^{\nu }}{s_{i,i-1}}-\frac{2 p_i^{\mu } p_{i-1}^{\nu }}{s_{i,i-1}}-\frac{2 p_j^{\mu } p_{j-1}^{\nu }}{s_{j,j-1}}-\frac{2 p_{j-1}^{\mu }
   p_j^{\nu }}{s_{j,j-1}}.
\eea
They satisfy separately our gauge conditions:
\bea
P_{ij}^\mu p_{i,\,\mu}&=&P_{ij}^\mu p_{i-1,\,\mu}=0\,,\nonumber\\
G_{ij}^{\mu\nu}p_{i,\mu}&=&G_{ij}^{\mu\nu}p_{i-1,\mu}=0^\nu\,,\nonumber\\
G_{ij}^{\mu\nu}p_{j,\nu}&=&G_{ij}^{\mu\nu}p_{j-1,\nu}=0^\mu.
\eea
We use the above definitions to define the 10 tensors as
\beq
\label{eq:gggg-tensor}
\vec{T}^{\mu\nu\rho\sigma}_{gggg}=\left(\begin{array}{c}G_{12}^{\mu\nu}G_{34}^{\rho\sigma}  \\ 
G_{13}^{\mu\rho}G_{24}^{\nu\sigma}\\
G_{14}^{\mu\sigma}G_{23}^{\nu\rho}\\
G_{12}^{\mu\nu} P^{\rho}_{31}P^{\sigma}_{42}\\
G_{13}^{\mu\rho} P^{\nu}_{24}P^{\sigma}_{42}\\
G_{14}^{\mu\sigma} P^{\nu}_{24}P^{\rho}_{31}\\
G_{23}^{\nu\rho} P^{\mu}_{13}P^{\sigma}_{42}\\
G_{24}^{\nu\sigma} P^{\mu}_{13}P^{\rho}_{31}\\
G_{34}^{\rho\sigma} P^{\mu}_{13}P^{\nu}_{24}\\
 P^{\mu}_{13}P^{\nu}_{24} P^{\rho}_{31}P^{\sigma}_{42}\\
\end{array}
\right).
\eeq

We decompose the four gluon scattering amplitude as
\bea
\label{eq:ggggdef}
\mathcal{A}_{gggg}&=&\epsilon^\mu_{a_1}(p_1)\epsilon^\nu_{a_2}(p_2)\epsilon^\rho_{a_3}(p_3)\epsilon^\sigma_{a_4}(p_4) \mathcal{A}_{gggg,\,\mu\nu\rho\sigma}^{a_1a_2a_3a_4}\nonumber\\
&=&\epsilon^\mu_{a_1}(p_1)\epsilon^\nu_{a_2}(p_2)\epsilon^\rho_{a_3}(p_3)\epsilon^\sigma_{a_4}(p_4) \sum_{i=1}^6 \sum_{j=1}^{10} c^{i,a_1a_2a_3a_4}_{gggg} T^{j}_{\mu\nu\rho\sigma,\,gggg} \mathcal{A}_{gggg}^{(i,j)}.
\eea
We refer to the coefficients $ \mathcal{A}_{gggg}^{(i,j)}$ as scalar amplitude coefficients.
Here, $\epsilon^\mu_{a}(p) $ is a polarisation vector with adjoint colour index $a$ corresponding to the gluon with momentum $p$.

Alternative to the representation of our amplitude introduced in eq.~\eqref{eq:ggggdef}, it is useful to decompose our amplitude into different components according to the helicities of the external gluons.
We can achieve this by projecting to two-dimensional external helicity states and by using the spinor-helicity framework (see for example ref.~\cite{Dixon:1996wi}).
Note, that the change of the dimension of external polarisation implies a change in the regularisation scheme used in the computation from CDR to HV scheme. 
The four-gluon scattering amplitude can then be written in this scheme as
\bea
\label{eq:gggg-hel}
\mathcal{A}_{gggg}&=&\epsilon_{a_1}\epsilon_{a_2}\epsilon_{a_3}\epsilon_{a_4} \sum_{i=1}^6 \sum_{j=1}^{16} c^{i,a_1a_2a_3a_4}_{gggg} h^{j}_{gggg} \mathcal{A}_{gggg}^{\prime\,\,(i,j)}.
\eea
In the above equation the $\epsilon^{a_i}$ now simply label the colour of an external gluon. 
The 16 helicity configurations are represented by the following structures,
\bea
\vec{h}_{gggg}&=&\Bigg\{\frac{[ 2  1 ] [ 4  3 ]}{\langle 1  2 \rangle \langle 3  4 \rangle},\frac{\langle 1  4 \rangle [ 2  1 ] [ 3  1 ]}{\langle 1  2 \rangle \langle 1  3 \rangle [ 4  1 ]},\frac{\langle 1  3 \rangle [ 2  1 ] [ 4  1 ]}{\langle 1  2 \rangle \langle 1  4 \rangle [ 3  1 ]},\frac{\langle 3  4 \rangle [ 2  1 ]}{\langle 1  2 \rangle [ 4  3 ]},\frac{\langle 2  3 \rangle [ 3  1 ] [ 4  3 ]}{\langle 1  3 \rangle \langle 3  4 \rangle [ 3  2 ]},\frac{\langle 1  4 \rangle \langle 2  3 \rangle [ 3  1 ]^2}{\langle 1  3 \rangle^2 [ 3  2 ] [ 4  1 ]},\nonumber\\
&&\frac{\langle 2  3 \rangle [ 4  1 ]}{\langle 1  4 \rangle [ 3  2 ]},\frac{\langle 2  3 \rangle \langle 3  4 \rangle [ 3  1 ]}{\langle 1  3 \rangle [ 3  2 ] [ 4  3 ]},\frac{\langle 1  3 \rangle [ 3  2 ] [ 4  3 ]}{\langle 2  3 \rangle \langle 3  4 \rangle [ 3  1 ]},\frac{\langle 1  4 \rangle [ 3  2 ]}{\langle 2  3 \rangle [ 4  1 ]},\frac{\langle 1  3 \rangle^2 [ 2  1 ] [ 4  3 ]}{\langle 1  2 \rangle \langle 3  4 \rangle [ 3  1 ]^2},\frac{\langle 1  3 \rangle \langle 1  4 \rangle [ 2  1 ]}{\langle 1  2 \rangle [ 3  1 ] [ 4  1 ]},\nonumber\\
&&\frac{\langle 1  2 \rangle [ 4  3 ]}{\langle 3  4 \rangle [ 2  1 ]},\frac{\langle 1  2 \rangle \langle 1  4 \rangle [ 3  1 ]}{\langle 1  3 \rangle [ 2  1 ] [ 4  1 ]},\frac{\langle 1  3 \rangle \langle 2  3 \rangle [ 4  3 ]}{\langle 3  4 \rangle [ 3  1 ] [ 3  2 ]},\frac{\langle 1  4 \rangle \langle 2  3 \rangle}{[ 3  2 ] [ 4  1 ]}\Bigg\}.
\eea
The above helicity structures correspond to positive (+) or negative (-) helicity as in
\bea
\vec{h}_{gggg} &\sim&\{{----},{---+},{--+-},{--++},{-+--},{-+-+},{-++-},{-+++},\nonumber\\
&&{+---},{+--+},{+-+-},{+-++},{++--},{++-+},{+++-},{++++}\}\,.
\eea
Here, the $i^{th}$ sign in each element of the list correspond to the helicity of the parton with momentum $p_i$.
The scalar coefficients of the amplitudes in the helicity basis can be written as a linear combination of the CDR amplitudes
\beq
\label{eq:helicityrel}
\mathcal{A}_{gggg}^{\prime\,\,(i,j)}=\sum\limits_{k=1}^{10} T_{CDR\rightarrow \text{helicity}, \,jk} \mathcal{A}_{gggg}^{(i,k)}.
\eeq
We provide the above transformation matrix $T_{CDR\rightarrow \text{helicity}} $ and $\mathcal{A}_{gggg}$ in electronic form together with the \arXiv~submission of this article. 
Note, that only 8 of the scalar helicity amplitudes are independent due to charge conjugation symmetry. The projection from CDR to HV is thus not invertible.

\section{The Two-Gluon and Two-Quark Amplitudes}
\label{sec:quark-gluon}
\noindent
\begin{figure}[h!]
\begin{center}
\includegraphics[width=0.30\textwidth]{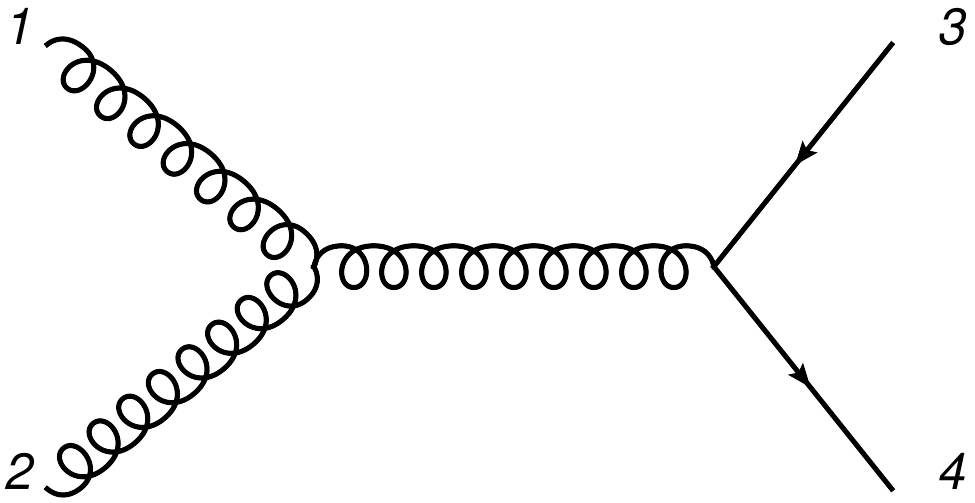}
\end{center}
\caption{Example of a leading order Feynman diagram for the scattering process of two gluons and two quarks.}
\label{fig:ggqq}
\end{figure}
The amplitude for the scattering of two gluons and two quarks can be decomposed into colour and Lorentz tensor structures.
The basis of colour structures for the amplitudes for the scattering of two gluons and two quarks can be brought in the form
\beq
\vec{c}_{ggq\bar q}=\left(\begin{array}{c} 
\delta^{a_1a_1}\delta_{i_4 i_3}\\
T^{a_2}_{i_4 i_5}T^{a_1}_{i_5 i_3}\\
T^{a_1}_{i_4 i_5}T^{a_2}_{i_5 i_3}\\
\end{array}\right).
\eeq
To derive the Lorentz tensor structures to represent our amplitude we begin with the following considerations.
In order to remove unphysical redundancies similar to the four gluon scattering case we work in a gauge where 
\beq
\epsilon(p_1).p_1=\epsilon(p_1).p_2=\epsilon(p_2).p_1=\epsilon(p_2).p_2=0.
\eeq
Owing to the momentum conservation we have $\bar u(p_3)\slashed p_2 u(p_4)=-\bar u(p_3)\slashed p_1 u(p_4)$.
Applying this identity and exploiting the restrictions imposed by the gauge choice we find that we can choose the following spinor and Lorentz structures as a basis to represent our amplitude
\beq
\label{eq:quarkgluontens}
\vec{T}^{\mu\nu}_{ggq\bar q}=\left(\begin{array}{c}
\Gamma^{\mu\nu}_\perp \\
\slashed p_1 g_\perp^{\mu\nu} \\
\slashed p_1 (P_3^\mu P_3^\nu-P_4^\mu P_3^\nu-P_4^\mu P_3^\nu+P_4^\mu P_4^\nu)\\
\gamma^\mu_\perp (P_3^\nu-P_4^\nu)\\
\gamma^\nu_\perp (P_3^\mu-P_4^\mu)
\end{array}\right).
\eeq
Here, we use the definitions 
\bea
\Gamma^{\mu\nu}_\perp&=&s_{12}\gamma^\mu \slashed p_1 \gamma^\nu +2 s_{13}\gamma^\nu p_1^\mu+2 s_{23}\gamma^\mu p_1^\mu \nonumber\\
&-& 4\slashed p_1 \left(p_1^\mu p_1^\nu+p_1^\mu p_3^\nu +p_1^\nu p_4^\nu \right)\,,\nonumber\\
g_\perp^{\mu\nu} &=&g^{\mu\nu}-\frac{2}{s_{12}}(p_1^\mu p_2^\nu+p_2^\mu p_1^\nu)\,,\nonumber\\
\gamma_\perp^{\mu} &=&\gamma^\mu+\slashed p_1 \frac{2}{s_{12}}(p_1^\mu-p_2^\mu)\,,\nonumber\\
P_i^{\mu} &=&s_{12}p_i^\mu-s_{2i}p_1^\mu-s_{1i}p_2^\mu\,. 
\eea
With this we may finally write the amplitude as
\beq
\label{eq:ggqqbardef}
\mathcal{A}_{ggq\bar q}= \epsilon_\mu^{a_1}(p_1)\epsilon_\nu^{a_2}(p_2) \sum\limits_{i=1}^3 \sum\limits_{j=1}^{5} \bar u_{b_4}(p_4) c^{i,\, a_1a_2}_{b_4b_3,\,ggq\bar q}T^{j,\,\mu\nu}_{ggq\bar q} u_{b_3}(p_3) \mathcal{A}^{(i,j)}_{ggq\bar q}.
\eeq

Furthermore, we may decompose our amplitude in components that correspond to different helicities of the external particles. 
In this decomposition, we write
\beq
\mathcal{A}_{ggq\bar q}= \epsilon^{a_1}\epsilon^{a_2}\bar v^{b_4} v^{b_3} \sum\limits_{i=1}^3 \sum\limits_{j=1}^{8}  c^{i,\, a_1a_2}_{b_4b_3,\,ggq\bar q}h^{j} _{ggq\bar q} \mathcal{A}^{\prime\,(i,j)}_{ggq\bar q}.
\eeq
The vectors $\bar v^{b_4}$ and $v^{b_3}$ carry the colour indices of the external anti-quark and quark.
The individual helicity structures are given by 
\bea
\label{eq:quarkgluonhel}
\vec{h}_{ggq\bar q}&=&\Big\{\frac{\langle 1  4 \rangle [ 2  1 ] [ 3  1 ]}{\langle 1  2 \rangle},\frac{\langle 1  3 \rangle [ 2  1 ] [ 4  1 ]}{\langle 1  2 \rangle},\frac{\langle 2  3 \rangle \langle 2  4 \rangle [ 3  1 ]}{\langle 1  3 \rangle},\frac{\langle 2  3 \rangle \langle 2  4 \rangle [ 4  1 ]}{\langle 1  4 \rangle},\frac{\langle 1  4 \rangle [ 3  2 ] [ 4  2 ]}{[ 4  1 ]},\nonumber\\
&&\frac{\langle 1  3 \rangle \langle 1  4 \rangle [ 4  2 ]}{\langle 2  4 \rangle},\frac{\langle 1  2 \rangle \langle 1  4 \rangle [ 3  1 ]}{[ 2  1 ]},\frac{\langle 1  2 \rangle \langle 1  3 \rangle [ 4  1 ]}{[ 2  1 ]}\Big\}.
\eea
The above helicity structures correspond to positive and negative helicities as in
\beq
\vec{h}_{ggq\bar q}\sim\{{---+},{--+-},{-+-+},{-++-},{+--+},{+-+-},{++-+},{+++-}\}.
\eeq
Here, the $i^{th}$ sign in each element of the list correspond to the helicity of the parton with momentum $p_i$.
Similarly to eq.~\eqref{eq:helicityrel} in the case of the four gluon amplitude, the scalar amplitude coefficients in the spinor-helicity representation can be written as a linear combination of the CDR amplitude coefficients.
We provide this linear relation in electronic form together with the \arXiv~submission of this article. 

Our scalar four particle amplitude is constructed such that if we consider the scattering kinematics of $p_1\, p_2\rightarrow p_3\, p_4$ all HPLs are real and the variable $x\in[0,1]$. 
Of course, this is not the only scattering process of a pair of quarks and gluons that can be considered. 
To obtain for example the amplitudes for the scattering of a quark and a gluon into a quark and a gluon we have to consider the kinematics of $p_1\, p_3\rightarrow p_2\,p_4$.
While the permutation of external momenta in the tensor structures in eq.~\eqref{eq:quarkgluontens} and eq.~\eqref{eq:quarkgluonhel} is trivial analytically continuing the HPLs in the scalar amplitudes is not.
How to perform this analytic continuation was discussed for example  in ref.~\cite{Henn:2019rgj} and we provide a brief summary in appendix~\ref{app:perm}.

\section{Four-Quark Amplitude}
\label{sec:quark}
\noindent
\begin{figure}[h!]
\begin{center}
\includegraphics[width=0.3\textwidth]{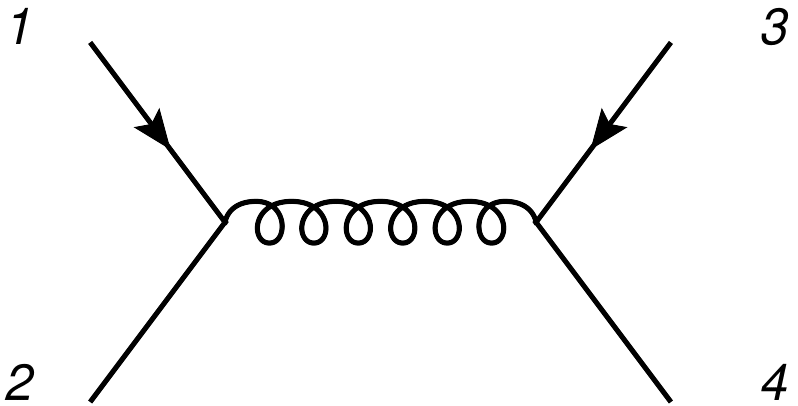}
\end{center}
\caption{Leading order Feynman diagram for the scattering process of four quarks.}
\label{fig:qqqq}
\end{figure}
\noindent
In the case of scattering amplitudes of four quarks we have to distinguish the case where all four quarks have the same flavour or two pairs of different flavours are scattering.
We refer to the scattering amplitude of the latter case as $\mathcal{A}_{q\bar q Q\bar Q}$. 
The former amplitude can then be expressed as 
\beq
\mathcal{A}_{q\bar q  q\bar q}(p_1,p_2,p_3,p_4)=\mathcal{A}_{q\bar q  Q\bar Q}(p_1,p_2,p_3,p_4)-\mathcal{A}_{q\bar q  Q\bar Q}(p_1,p_4,p_3,p_2).
\eeq
In the above equation we explicitly wrote the dependence of the amplitudes on the momenta to illustrate which momenta have to be exchanged in order to satisfy the relation.

The amplitude for the scattering of two quark- and anti-quark pairs of different flavours can be decomposed into colour and Lorentz structures as
\beq
\label{eq:qqbarQQbardef}
\mathcal{A}_{q\bar q Q \bar Q}=  \sum\limits_{i=1}^2 \sum\limits_{j=1}^{\infty} c^i_{q\bar q  Q\bar Q} T^j_{q\bar q  Q\bar Q}  \mathcal{A}^{(i,j)}_{q\bar q Q \bar Q}.
\eeq
In order to define the colour factors $c^i_{q\bar q  Q\bar Q}$ and spinor structures $T^j_{q\bar q  Q\bar Q}$ in the above equation, we find it convenient to decompose a spinor $u_i^a(p)$ with a fundamental colour index $a$ and spinor index $i$ into two factors
\beq
u_i^a(p)=u_i(p) v^a.
\eeq
The purpose of the vector $v^a$ is purely to label the colour of the external quark and $u_i(p)$ is a four component spinor.
The basis of colour structures for the amplitudes can be brought in the form
\beq
\vec{c}_{q\bar q  Q\bar Q}=\left(\begin{array}{c} 
\bar v^{a_1} \delta_{a_1 a_2}v^{a_2}\times \bar v^{a_3}\delta_{a_3a_4}v^{a_4}\\
\bar v^{a_1} \delta_{a_1 a_4}v^{a_4} \times \bar v^{a_3}\delta_{a_3a_2}v^{a_2}\\
\end{array}\right).
\eeq
The spinor structures can be written in terms of the following basis.
\beq
\label{eq:4qspinors}
\vec{T}_{q\bar q  Q\bar Q}=\left(\begin{array}{c}
\bar u(p_2) \slashed p_3 u(p_1) \times \bar u(p_4) \slashed p_1 u(p_3) \\
\bar u(p_2)  \gamma^\mu u(p_1) \times \bar u(p_4) \gamma_\mu u(p_3) \\
\bar u(p_2) \slashed p_3 \gamma^\mu\gamma^\nu u(p_1) \times \bar u(p_4) \slashed p_1 \gamma_\mu \gamma_\nu u(p_3) \\
\bar u(p_2)  \gamma^\mu\gamma^\nu\gamma^\rho u(p_1) \times \bar u(p_4)  \gamma_\mu \gamma_\nu\gamma_\rho u(p_3) \\
\bar u(p_2) \slashed p_3 \gamma^\mu\gamma^\nu \gamma^\rho \gamma^\sigma u(p_1) \times \bar u(p_4) \slashed p_1 \gamma_\mu\gamma_\nu\gamma_\rho \gamma_\sigma  u(p_3)  \\
\bar u(p_2) \gamma^\mu\gamma^\nu \gamma^\rho \gamma^\sigma \gamma^\tau u(p_1) \times \bar u(p_4) \gamma_\mu \gamma_\nu\gamma_\rho \gamma_\sigma\gamma_\tau  u(p_3) \\
\dots
\end{array}\right).
\eeq
Depending on the loop order we require more and more $\gamma$ matrices between the spinors. 
Specifically, to express the $L$- loop amplitude $2L-1$ $\gamma$-matrices are maximally required as soon as $L\geq 2$ (see for example ref.~\cite{Glover:2004si}).

Furthermore, we may decompose our amplitude in components that correspond to different helicities of the external particles.
We find 
\beq
\mathcal{A}_{q\bar q Q \bar Q}=  \sum\limits_{i=1}^2 \sum\limits_{j=1}^{4} c^i_{q\bar q  Q\bar Q} h^j_{q\bar q  Q\bar Q}  \mathcal{A}^{\prime\,(i,j)}_{q\bar q Q \bar Q}.
\eeq
Here the helicity basis is spanned by
\beq
\label{eq:4qhels}
\vec h_{q\bar q  Q\bar Q}  =\Big\{\frac{\langle 1  4 \rangle \langle 2  3 \rangle [ 3  1 ] }{\langle 1  3 \rangle},\langle 1  4 \rangle [ 3  2 ] ,\langle 2  3 \rangle [ 4  1 ] ,\frac{\langle 1  3 \rangle [ 2  1 ] [ 4  3 ]}{[ 3  1 ]} \Big\}.
\eeq
The above helicity structures correspond to positive and negative helicities as in
\beq
\vec h_{q\bar q  Q\bar Q}  \sim\{{-+-+},{-++-},{+--+},{+-+-}\}.
\eeq
Here, the $i^{th}$ sign in each element of the list correspond to the helicity of the parton with momentum $p_i$.
Similarly to eq.~\eqref{eq:helicityrel} in the case of the four gluon amplitude, the scalar amplitude coefficients in the spinor-helicity representation can be written as a linear combination of the CDR amplitude coefficients.
We provide this linear relation in electronic form together with the \arXiv~submission of this article. 

The variables in our amplitudes are chosen such that in the kinematic region of scattering particles with momenta $p_1$ and $p_2$ into particles with momenta $p_3$ and $p_4$ all HPLs are real and our variable $x\in[0,1]$.
However, this is just one particular case of scattering two quarks of each other and we may consider any permutation of momenta to cover all possible four quark scattering processes.
While obtaining such a permutation on the spinor structures of eq.~\eqref{eq:4qspinors} and eq.~\eqref{eq:4qhels} is relatively simple the analytic continuation of the HPLs in our scalar amplitude coefficients is non-trivial.
How such an analytic continuation can be obtained was discussed for example in ref.~\cite{Henn:2019rgj} and we outline it explicitly for convenience in appendix~\ref{app:perm}.

\section{Ultraviolet and Infrared Structure}
\label{sec:IR}
In this section, we review the structure of the poles in the dimensional regulator appearing in the four-point scattering amplitudes. 
The coefficients of the poles are well understood and can be expressed at least up to third order as products of known, universal factors and lower loop amplitudes.
Singularities arise due to soft (infrared) or collinear divergences of the loop integrands or due to ultraviolet (infinite momentum) divergences.
The possibility to derive the poles of an amplitude by direct computation serves as a stringent test of our computation. 
Not surprisingly we find agreement of our amplitudes with the structure of singularities outlined below.

Ultraviolet divergences in massless QCD scattering amplitudes can be universally reabsorbed into a renormalisation of the strong coupling constant. 
The renormalisation factor $Z$ can be derived by regarding the renormalisation group equation for $a_R$ (eq.~\eqref{eq:coup}):
\beq
\frac{\partial }{\partial \log(\mu^2)}a_R (\mu^2) =-\epsilon a_R(\mu^2)+a_R\beta.
\eeq
The $\beta$-function is related to the renormalisation factor $Z$ by
\beq
\beta=-\frac{\partial \log(Z)}{\partial \log (\mu^2)}=a_R \sum\limits_{i=0}^\infty a_R^{i}\beta_i.
\eeq
Solving the above evolution equation we can derive 
\beq
Z=1+\frac{a_R \beta _0}{\epsilon }+a_R^2 \left(\frac{\beta _0^2}{\epsilon
   ^2}+\frac{\beta _1}{2 \epsilon }\right) +a_R^3  \frac{\left(2 \beta _2 \epsilon ^2+7 \beta _1 \beta _0 \epsilon +6 \beta _0^3\right)}{6 \epsilon ^3}+\mathcal{O}(a_R^4).
\eeq
The QCD $\beta$-function coefficients are currently known to five-loop order~\cite{Herzog:2017,Baikov:2016}. 
We only require the first two for the purposes of this article which read
\bea
\beta_0&=&-\frac{1}{4}\left(\frac{11}{3}C_A-\frac{2}{3}n_f\right)\,,\nonumber\\
\beta_1&=&-\frac{1}{16}\left(\frac{34}{3}C_A^2-\frac{10}{3} C_A n_f-2 C_F n_f\right).
\eea
The quadratic Casimirs in adjoint and fundamental representations are denoted by $C_A=n_c$ and $C_F=(n_c^2-1)/2n_c$, respectively.

In general, a renormalised amplitude for the scattering of $n$ massless, SU($n_c)$ colour charged fields in dimensional regularisation can be written as 
\beq
\label{eq:ir-fac}
\mathcal{A}_n(\{p_i\},\epsilon)=\mathbf{Z}_n(\{p_i\},\epsilon)\mathcal{A}_n^f(\{p_i\},\epsilon)\,.
\eeq
Here $\mathcal{A}_n^f(\{p_i\})$ represents a finite hard amplitude depending on the the momenta $p_i$, colours and spins of the external states. 
The factor $\mathbf Z_n(\{p_i\},\epsilon)$ is universal and contains all infrared divergences~\cite{Catani:1998bh,Sterman:2002qn,Aybat:2006wq,Aybat:2006mz,Becher:2009cu,Gardi:2009qi,Almelid:2015jia,Almelid:2017qju}. 
To indicate an operator in colour space we use bold letters.
$\mathbf Z_n(\{p_i\},\epsilon)$ is given explicitly by the exponential
\beq
\label{eq:IRZ}
\mathbf{Z}_n(\{p_i\},\epsilon) =\mathcal{P} exp\left\{-\frac{1}{4}\int_0^{\mu^2} \frac{d\mu^2}{\mu^2} \mathbf \Gamma_n(\{p_i\},\mu^2,a_R(\mu^2)) \right\}.
\eeq
The soft anomalous dimension for $n$ fields is then given by
\beq
\label{eq:GammaDef}
\mathbf \Gamma_n(\{p_i\},\mu^2,a_R)=\mathbf \Gamma^{\text{dip.}}_n(\{p_i\},\mu^2,a_R)+\mathbf\Delta(\{\rho_{ijkl}\}).
\eeq
The first term on the right-hand side of the above equation corresponds to radiation from colour dipoles
\beq
\mathbf \Gamma^{\text{dip.}}_n(\{p_i\},\mu^2,a_R)=-\frac{1}{2}\gamma_c(a_R)\sum\limits_{i\neq j}^n\mathbf{T}_i\cdot\mathbf{T}_j \log\left(\frac{-s_{ij}}{\mu^2}\right) +\mathbf{\mathbb{I}} \sum\limits_{J=1}^n \gamma_J(a_R). 
\eeq
The operator $\mathbf T_i$ is commonly referred to as colour charge operator.
The action of such a colour charge operator on a scattering amplitude is given by regarding its action on individual spinors or polarisation vectors contracted with an amplitude 
\bea
\mathbf{T}^{a}_i \circ \epsilon^{b}_\mu(p_i) \mathcal{A}^\mu_b&=&-i f^{abc} \epsilon^\mu_b (p_i)\mathcal{A}_{\mu,\,c}\,,\nonumber\\
\mathbf{T}^{a}_i \circ  \mathcal{A}_{j,\,b} u_{j,\,b}(p_i)&=&\mathcal{A}_{j,\,b} T^a_{bc} u_{j,\,c}(p_i).
\eea
Here, $f^{abc}$ is SU($n_c$) structure constant and $T^a_{bc}$ is a fundamental generator. We introduce the abbreviation 
\beq
\mathbf{D}(\mu^2)=\sum\limits_{i> j=1}^n\mathbf{T}_i \cdot\mathbf{T}_j \log\left(\frac{-s_{ij}}{\mu^2}\right),\hspace{1cm} \mathbf{D}_0=\sum\limits_{i> j=1}^n\mathbf{T}_i\cdot\mathbf{T}_j.
\eeq
$\gamma_c$ is the cusp anomalous dimension and $\gamma_J$ are the anomalous dimensions associated with the external fields.
The second term on the right hand side of eq.~\eqref{eq:GammaDef} allows for soft radiation that relates more than two colour charged partons. 
The correction to the dipole formula~\cite{Almelid:2015jia,Almelid:2017qju} starts at third order in the coupling constant,
and will therefore not be needed in the present paper.

With this we now know the entire dependence of $\Gamma_n$ on the artificial scale $\mu^2$. 
Specifically, $\Gamma_n$ depends on $\mu^2$ only via a single logarithm and indirectly via $a_R$.
Carrying out the $\mu^2$-integral explicitly we find
\bea
&-&\frac{1}{2}\int_0^{\mu^2} \frac{d\mu^2}{\mu^2}  \mathbf \Gamma^{\text{dip.}}_n(\{p_i\},\mu^2,a_R(\mu^2))=a_R \left[\frac{\gamma_c^{(1)}}{8\epsilon^2}\mathbf{D}_0-\frac{\gamma_c^{(1)}}{8\epsilon}\mathbf{D}+\sum\limits_{J=1}^n\frac{\gamma_J^{(1)}}{2\epsilon}\mathbf{\mathbb{I}}\right]\\
&+&a_R^2 \left[\left( \frac{\beta_0\gamma_c^{(1)}}{32 \epsilon^3}+\frac{\gamma_c^{(2)}}{32 \epsilon^2}\right)\mathbf{D}_0+\left(-\frac{\beta_0\gamma_c^{(1)}}{16\epsilon^2}-\frac{\gamma_c^{(2)}}{16\epsilon}\right)\mathbf{D}+\sum\limits_{J=1}^n \left(\frac{\beta_0\gamma_J^{(1)}}{\epsilon^2}+\frac{\gamma_J^{(2)}}{4\epsilon}\right)\mathbf{\mathbb{I}}\right] +\mathcal{O}(a_R^3).\nonumber
\eea
Here, we expanded the anomalous dimensions in the strong coupling constant.
\beq
\gamma_c(a_R)=\sum\limits_{i=1}^\infty a_R^i\gamma_c^{(i)},\hspace{1cm}\gamma_J(a_R)=\sum\limits_{i=1}^\infty a_R^i\gamma_J^{(i)}.
\eeq
The first two terms in the expansion of the cusp anomalous dimensions are given by 
\bea
\gamma_c^{(1)}&=&1\,,\nonumber\\
\gamma_c^{(2)}&=&C_A \left( \frac{67 }{36}-\frac{1}{2} \zeta_2 \right)-\frac{5 }{18}n_f.
\eea
The first two terms in the expansion of the collinear anomalous dimensions are given by
\bea
\gamma_g^{(1)}&=&-\frac{11}{12}C_A+\frac{1}{6}n_f,\nonumber\\
\gamma_q^{(1)}&=&-\frac{3}{4}C_F,\nonumber\\
\gamma_g^{(2)}&=& C_A^2 \left(-\frac{173}{108} +\frac{11 \pi ^2}{288} +\frac{\zeta_3}{8} \right)+C_A n_f \left(\frac{8}{27}-\frac{\pi^2}{144} \right)+\frac{1}{8}C_F n_f\,,\nonumber\\
\gamma_q^{(2)}&=&C_A C_F \Big( -\frac{961}{864}-\frac{11\pi^2}{96}+\frac{13 \zeta_3}{8}\Big)+ C_F^2 \Big( -\frac{3}{32}+\frac{\pi^2}{8}-\frac{3\zeta_3}{2}\Big) \nonumber\\
&&+ C_F n_f \Big( \frac{65}{432}+\frac{\pi^2}{48}\Big)\,.
\eea

\section{Checks on Results}
\label{sec:checks}
We perform a number of consistency checks on the scattering amplitudes to ensure the correctness of our result.
\begin{itemize}
\item As described in section \ref{sec:setup}, all amplitudes were generated by two independent calculational setups.
    \item All four-point scattering amplitudes exhibit the universal infrared structures predicted by eq.~(\ref{eq:IRZ}). This serves as the most stringent check on our results. 
    \item The four-gluon helicity amplitudes are consistent with the existing results computed in ref.~\cite{Bern:2002tk} to ${\cal O}(\epsilon^0)$. 
    Moreover, interfering the CDR amplitude with its tree level equivalent reproduces correctly the results of ref.~\cite{Glover:2001af}.
    \item We explicitly checked that only 4 helicity amplitudes ($++++, -+++, --++, -+-+$) for four gluon scattering are independent (as dictated by Bose, parity and time-reversal symmetry). 
    \item The one-loop U(1) decoupling identity~\cite{Bern:1990ux} for the four-gluon amplitude is satisfied. 
    This identity states that the double trace coefficients of the pure gauge part of the four-gluon amplitude for any helicity configuration should be related to the corresponding single trace elements through
    \begin{align}
        {\cal A}_{gggg}^{\prime  (4,j,1)} = {\cal A}_{gggg}^{\prime  (5,j,1)} = {\cal A}_{gggg}^{\prime  (6,j,1)} = \frac{2}{n_c}\Big( {\cal A}_{gggg}^{\prime  (1,j,1)}+{\cal A}_{gggg}^{\prime  (2,j,1)}+{\cal A}_{gggg}^{\prime  (3,j,1)}\Big)\,.
    \end{align}
    In the above equation, the quantity ${\cal A}_{gggg}^{\prime\, (i,j,k)}$ is defined through
\begin{align}
    {\cal A}_{gggg}^{\prime\,(i,j)} = \frac{\alpha}{4\pi}\sum_{k=0}^{\infty} \left(\frac{\alpha}{4\pi}\right)^k {\cal A}_{gggg}^{\prime\,(i,j,k)}\,,
\end{align}
where ${\cal A}_{gggg}^{\prime\,(i,j)}$ is introduced in eq.~(\ref{eq:gggg-hel}).

Moreover, in ref.~\cite{Naculich:2011ep}, more general group theory constraints are discussed which are also verified for the two-loop helicity amplitudes. In order to present those, we rewrite the $n_f$ independent part of the four gluon amplitude in eq.~(\ref{eq:gggg-hel}) as
    \begin{align}
    \label{eq:gggg-hel-alt-decom}
    \mathcal{A}_{gggg}=&\epsilon_{a_1}\epsilon_{a_2}\epsilon_{a_3}\epsilon_{a_4} \sum_{i=1}^3 \sum_{j=1}^{16} \sum_{k=0}^{\infty} \sum_{m=0}^{\lfloor \frac{k}{2} \rfloor} n_c^{k-2m} c^{i,a_1a_2a_3a_4}_{gggg} h^{j}_{gggg} \left(\frac{\alpha_s}{4\pi}\right)^{k+1}  \mathcal{A}_{gggg}^{\prime\,(i,j,k,2m)}\,\nonumber\\
    +&\epsilon_{a_1}\epsilon_{a_2}\epsilon_{a_3}\epsilon_{a_4} \sum_{i=4}^6 \sum_{j=1}^{16} \sum_{k=0}^{\infty} \sum_{m=0}^{\lfloor \frac{k-1}{2} \rfloor} n_c^{k-2m-1} c^{i,a_1a_2a_3a_4}_{gggg} h^{j}_{gggg} \left(\frac{\alpha_s}{4\pi}\right)^{k+1}  \mathcal{A}_{gggg}^{\prime\,(i,j,k,2m+1)}\,.
    \end{align}
   The constraints take the form 
   \begin{align}
   &\sum_{i=1}^3 {\cal A}_{gggg}^{\prime  (i,j,2,2)}=0\,,\nonumber\\
   &6\sum_{i=1}^3 {\cal A}_{gggg}^{\prime  (i,j,2,0)} - \sum_{i=4}^6 {\cal A}_{gggg}^{\prime  (i,j,2,1)}=0\,,\nonumber\\
   &{\cal A}_{gggg}^{\prime  (i,j,2,2)} + {\cal A}_{gggg}^{\prime  ((i+4)~ {\rm modulo}~ 6,j,2,1)}={\text{independent of }} i\,.
   \end{align}
    \item We find complete agreement with ref.~\cite{Bern:2003ck} for the one- and two-loop helicity amplitudes for the two-gluon and two-quark channel.   
    \item We find that one- and two-loop helicity amplitudes for four-quark scattering with two different flavours are in agreement with the existing results computed in ref.~\cite{DeFreitas:2004kmi}. 
    
Note that in the aforementioned references~\cite{Bern:2002tk,Bern:2003ck,DeFreitas:2004kmi}, the finite remainder is defined through subtraction operators which are different from the ones used in this article at ${\cal O}(\epsilon^0)$. The difference is taken into account while performing the comparison.
\end{itemize}
\section{Conclusions}
\label{sec:concl}

As a main result of this article, we provide the scalar amplitude coefficients of the four-particle scattering amplitudes in CDR together with the \arXiv~submission of the article. 
We include the LO, NLO and NNLO coefficients computed to sixth, fourth and second power in the Laurent expansion in the dimensional regulator respectively in machine readable form. 
Furthermore, we include the transformation matrices $T_{CDR\rightarrow \text{helicity}}$ to easily extract the amplitudes coefficients in the spinor-helicity basis.

Our result represents the first step towards a computation of three-loop, four-parton scattering amplitudes in QCD.
In particular, our results can be used to predict the infrared and collinear pole structure of such three loop amplitudes.
Furthermore, they are a necessary ingredient to obtain a suitable ultraviolet, infrared and collinear subtraction term for future cross section computations. 

Terms beyond the finite order in the dimensional regulator are obtained in this article for the first time.
Beyond their application to future collider phenomenology these terms may serve as analytic data for the analysis of QCD scattering amplitudes, such as the Regge limit.
Our amplitudes are represented in the form of widely used harmonic polylogarithms in a uniform notation which makes their application and analysis simple.

With this work we have taken a very first step towards the formidable task of computing one of the most crucial collider processes - the production of two hadronic jets - to third order in QCD perturbation theory.

\section*{Acknowledgements}
B.~M. thanks E. Gardi and V. Del Duca for discussions about the high energy limit. T.~A. thanks J. Davies, M.~K. Mandal, V. Ravindran and M. Steinhauser for various discussions. This research received funding from the European Research Council (ERC) under the European Union's Horizon 2020 research and innovation programme (grant agreement No 725110), \textit{Novel structures
in scattering amplitudes}.
B.~M. is supported by the Pappalardo fellowship.

\appendix
\section{Permutation of External Momenta}
\label{app:perm}

The results for the four point scattering amplitude of this paper are presented such that in the particular kinematic region of particles with momenta $p_1$ and $p_2$ scattering into particles $p_3$ and $p_4$ all HPLs are real valued and the evaluation of our amplitudes is straight forward.
However, this particular scattering region is not the only scattering region of interest.
For example, relating our amplitudes to the case of particles with momentum $p_1$ and $p_3$ scattering into particles with momentum $p_2$ and $p_4$ is non-trivial.
Consequently, it is useful to describe how such a relation can be achieved. 
Mainly we have to concern ourselves with the transformation of HPLs from one scattering region to the other. 
How this can be achieved was described for example in ref.~\cite{Henn:2019rgj} but we find it convenient to outline the basic steps here.

First, it is prudent to recall that all our L-loop amplitudes have a pre-factor 
\beq
\left(\frac{s_{12}}{\mu^2}\right)^{-L \epsilon} =1-L\epsilon \log\left(\frac{s_{12}}{\mu^2}\right)+\mathcal{O}(\epsilon^2).
\eeq
Naturally, this pre-factor needs to be taken into account when permutaions of external legs are performed. 
Furthermore, it is necessary to restore the correct energy dimension of the scalar amplitudes since we chose for convenience $s_{12}=1$ in our results.
While the permutation of external legs is obvious for tensor and helicity structures, this is not the case for the analytic functions appearing in our amplitudes.
Let us first consider the permutation where momentum $p_1$ and $p_2$ are exchanged. 
This leaves the Mandelstam invariant $s_{12}$ invariant and exchanges $s_{13}$ and $s_{23}$ and replaces $x\rightarrow 1-x$. 
In the scattering region we used to define our amplitudes both $s_{13}$ and $s_{23}$ are negative and no branch cut of our amplitude is crossed by performing this permutation.
Transforming the HPLs can be done by standard methods (see for example ref.~\cite{Maitre:2005uu}) and does not introduce new imaginary parts.

If we consider scattering kinematics where a Mandelstam invariant changes its sign (for example $s_{12}<0$) we need to analytically continue the HPLs of our amplitudes. 
To achieve this we may attribute an infinitesimally small imaginary part to each Lorentz invariant scalar product of momenta:
\bea
s_{12}&\rightarrow &s_{12}-i 0\,,\\
s_{13}&\rightarrow &s_{13}-i 0\,,\nonumber\\
s_{23}&\rightarrow &s_{23}-i 0\,.\nonumber
\eea
However, in our calculation we applied the relation $s_{12}+s_{13}+s_{23}=0$ which arises due to momentum conservation. 
As a consequence the above assignment cannot be done uniquely and it is ambiguous which infinitesimal phase should be attributed to the variable x.
For example, consider two different possibilities of thinking about the polynomial $1-x$.
\begin{enumerate}
\item 
\beq
1-x=-\frac{s_{23}-i 0}{s_{12}-i0}=-\frac{s_{23}}{s_{12}}+i 0.
\eeq
\item 
\beq
1-x=1+\frac{s_{13}-i 0}{s_{12}-i0}=1+\frac{s_{13}}{s_{12}}-i 0.
\eeq
\end{enumerate}
In the above equations the sign of the infinitesimal imaginary part differs, which demonstrates the problematic ambiguity.
To perform an analytic continuation from one scattering region to another after having applied the momentum conservation identity we need to add another bit of information.
The branch cuts of our scattering amplitude are located at kinematic points where Mandelstam invariants vanish.
If we can manifest the branch cuts of our amplitude in terms of explicit logarithms we can easily identify the argument of these logarithms as being ratios of Mandelstam invariants.
We discuss how this can be achieved at hands of a specific example below.

We consider the permutation of momenta $p_2\leftrightarrow p_3$. We find that our variable $x$ changes:
\beq
x=-\frac{(p_1+p_3)^2}{(p_1+p_2)^2}\rightarrow -\frac{(p_1+p_2)^2}{(p_1+p_3)^2}=\frac{1}{y}\,.
\eeq
In the new, desired scattering region $p_1\,p_3 \rightarrow p_2\, p_4$ we find $y\in[0,1]$.
However, we need to be careful when transforming the logarithmic branch cuts of our amplitude.
\bea
\log(s_{12}-i0)\rightarrow\log(s_{13}-i0)&=&\log(s_{12}-i0) +\log\left(\frac{-s_{13}+i0}{s_{12}-i0}\right) -i\pi\nonumber\\
&=&\log(s_{12}-i0) +\log\left(y\right) -i\pi.\nonumber\\
\log(x)=\log\left(\frac{-s_{13}+i0}{s_{12}-i0}\right)&\rightarrow&-\log(y)+2 i\pi.\nonumber\\
\log(1-x)=\log\left(\frac{-s_{23}+i0}{s_{12}-i0}\right)&\rightarrow&\log(1-y)-\log(y)+i\pi.
\eea
All branch points of our amplitudes are located at $x=0$, $x=1$ and $x=\infty$.
As we can see from the above equations all three branch cuts are crossed when the desired permutation is performed. 
This complicates the transformation of HPLs and we discuss our solution in the following. 
Our amplitudes are comprised of HPLs  and consequently only one branch cut can be made manifest in terms of explicit logarithms at a time.
In particular, this can be achieved by shuffle identities of the HPL's such as
\beq
H(a_n,\dots,a_2,0,x)=H(a_n,\dots,a_2,x)\log(x)-H(0,a_n,\dots,a_2,x)-\dots-H(a_n,\dots,0,a_2,x)
\eeq
or generalisations thereof, that make branch cuts (here at $x=0$) explicit.
In order to perform the permutation corresponding to the exchange of momenta $p_2\leftrightarrow p_3$ we proceed in three sequential steps.
In each of the steps only one branch cut at $x=0$  or $x=1$ is crossed. 
This implies that we may use shuffle identities to make the branch cut in our amplitude explicit and that the transformation of the remaining HPLs at each step does not introduce any imaginary parts.
The three steps are given by the following.
\begin{enumerate}
\item Analytically continue into a scattering region where $s_{12}^\prime>0$, $s_{13}^\prime>0$ and $s_{23}^\prime<0$ and define $x^\prime=-x$.
\bea
\log(s_{12}-i0)&\rightarrow & \log(s_{12}^\prime-i0).\nonumber\\
\log(x)&\rightarrow & \log(x^\prime)+i\pi.\nonumber\\
\log(1-x)&\rightarrow & \log(1+x^\prime).
\eea
\item Permute the external momenta $p_2\leftrightarrow p_3$ and introduce $x^{\prime \prime}=\frac{1}{x^\prime}$. 
\bea
\log(s_{12}^\prime-i0)&\rightarrow &\log(s_{12}^{\prime\prime}-i0) +\log\left(x^{\prime\prime}\right).\nonumber\\
\log(x^\prime)&\rightarrow & -\log(x^{\prime\prime}).\nonumber\\
\log(1+x^\prime)&\rightarrow & \log(1+x^{\prime\prime})-\log(x^{\prime\prime}).\nonumber\\
\eea
\item Analytically continue into a scattering region where $s_{12}^{\prime\prime\prime}>0$, $s_{13}^{\prime\prime\prime}<0$ and $s_{23}^{\prime\prime\prime}<0$ and define $x^{\prime\prime\prime}=-x^{\prime\prime}$.
\bea
\log(s_{12}^{\prime\prime}-i0)&\rightarrow & \log(s_{12}^{\prime\prime\prime}-i0).\nonumber\\
\log(x^{\prime\prime})&\rightarrow & \log(x^{\prime\prime\prime})-i\pi.\nonumber\\
\log(1+x^{\prime\prime})&\rightarrow & \log(1-x^{\prime\prime\prime}).
\eea
\end{enumerate}
Having completed the three transformations we performed the desired permutation and may drop the ${}^\prime$ labels for convenience.

Finally, we note that any other permutation of external legs can simply be a sequence of the two example permutations outlined above. 
With this we showed how to obtain the four-point scattering amplitude for any scattering kinematics starting from the amplitudes provided in this article.

\bibliographystyle{JHEP}
\bibliography{main}

\providecommand{\href}[2]{#2}\begingroup\raggedright\begin{thebibliography}{10}

\bibitem{Aaboud:2017wsi}
{\scshape ATLAS} collaboration, \emph{{Measurement of inclusive jet and dijet
  cross-sections in proton-proton collisions at $\sqrt{s}=13$ TeV with the
  ATLAS detector}}, \href{https://doi.org/10.1007/JHEP05(2018)195}{\emph{JHEP}
  {\bfseries 05} (2018) 195}
  [\href{https://arxiv.org/abs/1711.02692}{{\ttfamily 1711.02692}}].

\bibitem{Khachatryan:2016wdh}
{\scshape CMS} collaboration, \emph{{Measurement of the double-differential
  inclusive jet cross section in proton–proton collisions at $\sqrt{s} =
  13\,\text {TeV} $}},
  \href{https://doi.org/10.1140/epjc/s10052-016-4286-3}{\emph{Eur. Phys. J.}
  {\bfseries C76} (2016) 451}
  [\href{https://arxiv.org/abs/1605.04436}{{\ttfamily 1605.04436}}].

\bibitem{Sirunyan:2017skj}
{\scshape CMS} collaboration, \emph{{Measurement of the triple-differential
  dijet cross section in proton-proton collisions at $\sqrt{s}=8\,\text {TeV} $
  and constraints on parton distribution functions}},
  \href{https://doi.org/10.1140/epjc/s10052-017-5286-7}{\emph{Eur. Phys. J.}
  {\bfseries C77} (2017) 746}
  [\href{https://arxiv.org/abs/1705.02628}{{\ttfamily 1705.02628}}].

\bibitem{Gao:2012he}
J.~Gao, Z.~Liang, D.~E. Soper, H.-L. Lai, P.~M. Nadolsky and C.~P. Yuan,
  \emph{{MEKS: a program for computation of inclusive jet cross sections at
  hadron colliders}},
  \href{https://doi.org/10.1016/j.cpc.2013.01.022}{\emph{Comput. Phys. Commun.}
  {\bfseries 184} (2013) 1626}
  [\href{https://arxiv.org/abs/1207.0513}{{\ttfamily 1207.0513}}].

\bibitem{Alioli:2010xa}
S.~Alioli, K.~Hamilton, P.~Nason, C.~Oleari and E.~Re, \emph{{Jet pair
  production in POWHEG}},
  \href{https://doi.org/10.1007/JHEP04(2011)081}{\emph{JHEP} {\bfseries 04}
  (2011) 081} [\href{https://arxiv.org/abs/1012.3380}{{\ttfamily 1012.3380}}].

\bibitem{Giele:1994gf}
W.~T. Giele, E.~W.~N. Glover and D.~A. Kosower, \emph{{The Two-Jet Differential
  Cross Section at ${\cal O}(\alpha_s^3)$ in Hadron Collisions}},
  \href{https://doi.org/10.1103/PhysRevLett.73.2019}{\emph{Phys. Rev. Lett.}
  {\bfseries 73} (1994) 2019}
  [\href{https://arxiv.org/abs/hep-ph/9403347}{{\ttfamily hep-ph/9403347}}].

\bibitem{Ellis:1992en}
S.~D. Ellis, Z.~Kunszt and D.~E. Soper, \emph{{Two jet production in hadron
  collisions at order alpha-s**3 in QCD}},
  \href{https://doi.org/10.1103/PhysRevLett.69.1496}{\emph{Phys. Rev. Lett.}
  {\bfseries 69} (1992) 1496}.

\bibitem{Currie:2017eqf}
J.~Currie, A.~Gehrmann-De~Ridder, T.~Gehrmann, E.~W.~N. Glover, A.~Huss and
  J.~Pires, \emph{{Precise predictions for dijet production at the LHC}},
  \href{https://doi.org/10.1103/PhysRevLett.119.152001}{\emph{Phys. Rev. Lett.}
  {\bfseries 119} (2017) 152001}
  [\href{https://arxiv.org/abs/1705.10271}{{\ttfamily 1705.10271}}].

\bibitem{Gehrmann-DeRidder:2019ibf}
A.~Gehrmann-De~Ridder, T.~Gehrmann, E.~W.~N. Glover, A.~Huss and J.~Pires,
  \emph{{Triple Differential Dijet Cross Section at the LHC}},
  \href{https://doi.org/10.1103/PhysRevLett.123.102001}{\emph{Phys. Rev. Lett.}
  {\bfseries 123} (2019) 102001}
  [\href{https://arxiv.org/abs/1905.09047}{{\ttfamily 1905.09047}}].

\bibitem{Currie:2018oxh}
J.~Currie, A.~Gehrmann-De~Ridder, T.~Gehrmann, N.~Glover, A.~Huss and J.~Pires,
  \emph{{Jet cross sections at the LHC with NNLOJET}},
  \href{https://doi.org/10.22323/1.303.0001}{\emph{PoS} {\bfseries LL2018}
  (2018) 001} [\href{https://arxiv.org/abs/1807.06057}{{\ttfamily
  1807.06057}}].

\bibitem{Czakon:2019tmo}
M.~Czakon, A.~van Hameren, A.~Mitov and R.~Poncelet, \emph{{Single-jet
  inclusive rates with exact color at $\mathcal{O}(\alpha_s^4)$}},
  \href{https://arxiv.org/abs/1907.12911}{{\ttfamily 1907.12911}}.

\bibitem{Anastasiou:2000kg}
C.~Anastasiou, E.~W.~N. Glover, C.~Oleari and M.~E. Tejeda-Yeomans,
  \emph{{Two-loop QCD corrections to the scattering of massless distinct
  quarks}}, \href{https://doi.org/10.1016/S0550-3213(01)00079-7}{\emph{Nucl.
  Phys.} {\bfseries B601} (2001) 318}
  [\href{https://arxiv.org/abs/hep-ph/0010212}{{\ttfamily hep-ph/0010212}}].

\bibitem{Anastasiou:2000ue}
C.~Anastasiou, E.~W.~N. Glover, C.~Oleari and M.~E. Tejeda-Yeomans, \emph{{Two
  loop QCD corrections to massless identical quark scattering}},
  \href{https://doi.org/10.1016/S0550-3213(01)00080-3}{\emph{Nucl. Phys.}
  {\bfseries B601} (2001) 341}
  [\href{https://arxiv.org/abs/hep-ph/0011094}{{\ttfamily hep-ph/0011094}}].

\bibitem{Bern:2000dn}
Z.~Bern, L.~J. Dixon and D.~A. Kosower, \emph{{A Two loop four gluon helicity
  amplitude in QCD}},
  \href{https://doi.org/10.1088/1126-6708/2000/01/027}{\emph{JHEP} {\bfseries
  01} (2000) 027} [\href{https://arxiv.org/abs/hep-ph/0001001}{{\ttfamily
  hep-ph/0001001}}].

\bibitem{Glover:2001af}
E.~W.~N. Glover, C.~Oleari and M.~E. Tejeda-Yeomans, \emph{{Two loop QCD
  corrections to gluon-gluon scattering}},
  \href{https://doi.org/10.1016/S0550-3213(01)00210-3}{\emph{Nucl. Phys.}
  {\bfseries B605} (2001) 467}
  [\href{https://arxiv.org/abs/hep-ph/0102201}{{\ttfamily hep-ph/0102201}}].

\bibitem{Anastasiou:2001sv}
C.~Anastasiou, E.~W.~N. Glover, C.~Oleari and M.~E. Tejeda-Yeomans, \emph{{Two
  loop QCD corrections to massless quark gluon scattering}},
  \href{https://doi.org/10.1016/S0550-3213(01)00195-X}{\emph{Nucl. Phys.}
  {\bfseries B605} (2001) 486}
  [\href{https://arxiv.org/abs/hep-ph/0101304}{{\ttfamily hep-ph/0101304}}].

\bibitem{Bern:2002tk}
Z.~Bern, A.~De~Freitas and L.~J. Dixon, \emph{{Two loop helicity amplitudes for
  gluon-gluon scattering in QCD and supersymmetric Yang-Mills theory}},
  \href{https://doi.org/10.1088/1126-6708/2002/03/018}{\emph{JHEP} {\bfseries
  03} (2002) 018} [\href{https://arxiv.org/abs/hep-ph/0201161}{{\ttfamily
  hep-ph/0201161}}].

\bibitem{Bern:2003ck}
Z.~Bern, A.~De~Freitas and L.~J. Dixon, \emph{{Two loop helicity amplitudes for
  quark gluon scattering in QCD and gluino gluon scattering in supersymmetric
  Yang-Mills theory}}, \href{https://doi.org/10.1007/JHEP04(2014)112,
  10.1088/1126-6708/2003/06/028}{\emph{JHEP} {\bfseries 06} (2003) 028}
  [\href{https://arxiv.org/abs/hep-ph/0304168}{{\ttfamily hep-ph/0304168}}].

\bibitem{DeFreitas:2004kmi}
A.~De~Freitas and Z.~Bern, \emph{{Two-loop helicity amplitudes for quark-quark
  scattering in QCD and gluino-gluino scattering in supersymmetric Yang-Mills
  theory}}, \href{https://doi.org/10.1088/1126-6708/2004/09/039}{\emph{JHEP}
  {\bfseries 09} (2004) 039}
  [\href{https://arxiv.org/abs/hep-ph/0409007}{{\ttfamily hep-ph/0409007}}].

\bibitem{DeFreitas:2004aj}
A.~De~Freitas and Z.~Bern, \emph{{Two-loop helicity amplitudes for
  fermion-fermion scattering}},
  \href{https://doi.org/10.1016/j.nuclphysbps.2004.09.010}{\emph{Nucl. Phys.
  Proc. Suppl.} {\bfseries 135} (2004) 51}
  [\href{https://arxiv.org/abs/hep-ph/0409036}{{\ttfamily hep-ph/0409036}}].

\bibitem{Abreu:2017xsl}
S.~Abreu, F.~Febres~Cordero, H.~Ita, M.~Jaquier, B.~Page and M.~Zeng,
  \emph{{Two-Loop Four-Gluon Amplitudes from Numerical Unitarity}},
  \href{https://doi.org/10.1103/PhysRevLett.119.142001}{\emph{Phys. Rev. Lett.}
  {\bfseries 119} (2017) 142001}
  [\href{https://arxiv.org/abs/1703.05273}{{\ttfamily 1703.05273}}].

\bibitem{Glover:2003cm}
E.~W.~N. Glover and M.~E. Tejeda-Yeomans, \emph{{Two loop QCD helicity
  amplitudes for massless quark massless gauge boson scattering}},
  \href{https://doi.org/10.1088/1126-6708/2003/06/033}{\emph{JHEP} {\bfseries
  06} (2003) 033} [\href{https://arxiv.org/abs/hep-ph/0304169}{{\ttfamily
  hep-ph/0304169}}].

\bibitem{Glover:2004si}
E.~W.~N. Glover, \emph{{Two loop QCD helicity amplitudes for massless quark
  quark scattering}},
  \href{https://doi.org/10.1088/1126-6708/2004/04/021}{\emph{JHEP} {\bfseries
  04} (2004) 021} [\href{https://arxiv.org/abs/hep-ph/0401119}{{\ttfamily
  hep-ph/0401119}}].

\bibitem{Broggio:2014hoa}
A.~Broggio, A.~Ferroglia, B.~D. Pecjak and Z.~Zhang, \emph{{NNLO hard functions
  in massless QCD}}, \href{https://doi.org/10.1007/JHEP12(2014)005}{\emph{JHEP}
  {\bfseries 12} (2014) 005} [\href{https://arxiv.org/abs/1409.5294}{{\ttfamily
  1409.5294}}].

\bibitem{Luo2019}
Q.~Jin and H.~Luo, \emph{{Analytic Form of the Three-loop Four-gluon Scattering
  Amplitudes in Yang-Mills Theory}},
  \href{https://arxiv.org/abs/1910.05889}{{\ttfamily 1910.05889}}.

\bibitem{tHooft:1972tcz}
G.~'t~Hooft and M.~J.~G. Veltman, \emph{{Regularization and Renormalization of
  Gauge Fields}},
  \href{https://doi.org/10.1016/0550-3213(72)90279-9}{\emph{Nucl. Phys.}
  {\bfseries B44} (1972) 189}.

\bibitem{Bern:1991aq}
Z.~Bern and D.~A. Kosower, \emph{{The Computation of loop amplitudes in gauge
  theories}}, \href{https://doi.org/10.1016/0550-3213(92)90134-W}{\emph{Nucl.
  Phys.} {\bfseries B379} (1992) 451}.

\bibitem{Berends:1981rb}
F.~A. Berends, R.~Kleiss, P.~De~Causmaecker, R.~Gastmans and T.~T. Wu,
  \emph{{Single Bremsstrahlung Processes in Gauge Theories}},
  \href{https://doi.org/10.1016/0370-2693(81)90685-7}{\emph{Phys. Lett.}
  {\bfseries 103B} (1981) 124}.

\bibitem{DeCausmaecker:1981wzb}
P.~De~Causmaecker, R.~Gastmans, W.~Troost and T.~T. Wu, \emph{{Helicity
  Amplitudes for Massless QED}},
  \href{https://doi.org/10.1016/0370-2693(81)91025-X}{\emph{Phys. Lett.}
  {\bfseries 105B} (1981) 215}.

\bibitem{Xu:1986xb}
Z.~Xu, D.-H. Zhang and L.~Chang, \emph{{Helicity Amplitudes for Multiple
  Bremsstrahlung in Massless Nonabelian Gauge Theories}},
  \href{https://doi.org/10.1016/0550-3213(87)90479-2}{\emph{Nucl. Phys.}
  {\bfseries B291} (1987) 392}.

\bibitem{Broggio:2015dga}
A.~Broggio, C.~Gnendiger, A.~Signer, D.~Stöckinger and A.~Visconti,
  \emph{{SCET approach to regularization-scheme dependence of QCD amplitudes}},
  \href{https://doi.org/10.1007/JHEP01(2016)078}{\emph{JHEP} {\bfseries 01}
  (2016) 078} [\href{https://arxiv.org/abs/1506.05301}{{\ttfamily
  1506.05301}}].

\bibitem{Catani:1998bh}
S.~Catani, \emph{{The Singular behavior of QCD amplitudes at two loop order}},
  \href{https://doi.org/10.1016/S0370-2693(98)00332-3}{\emph{Phys. Lett.}
  {\bfseries B427} (1998) 161}
  [\href{https://arxiv.org/abs/hep-ph/9802439}{{\ttfamily hep-ph/9802439}}].

\bibitem{Sterman:2002qn}
G.~F. Sterman and M.~E. Tejeda-Yeomans, \emph{{Multiloop amplitudes and
  resummation}},
  \href{https://doi.org/10.1016/S0370-2693(02)03100-3}{\emph{Phys. Lett.}
  {\bfseries B552} (2003) 48}
  [\href{https://arxiv.org/abs/hep-ph/0210130}{{\ttfamily hep-ph/0210130}}].

\bibitem{Aybat:2006wq}
S.~M. Aybat, L.~J. Dixon and G.~F. Sterman, \emph{{The Two-loop anomalous
  dimension matrix for soft gluon exchange}},
  \href{https://doi.org/10.1103/PhysRevLett.97.072001}{\emph{Phys. Rev. Lett.}
  {\bfseries 97} (2006) 072001}
  [\href{https://arxiv.org/abs/hep-ph/0606254}{{\ttfamily hep-ph/0606254}}].

\bibitem{Aybat:2006mz}
S.~M. Aybat, L.~J. Dixon and G.~F. Sterman, \emph{{The Two-loop soft anomalous
  dimension matrix and resummation at next-to-next-to leading pole}},
  \href{https://doi.org/10.1103/PhysRevD.74.074004}{\emph{Phys. Rev.}
  {\bfseries D74} (2006) 074004}
  [\href{https://arxiv.org/abs/hep-ph/0607309}{{\ttfamily hep-ph/0607309}}].

\bibitem{Becher:2009cu}
T.~Becher and M.~Neubert, \emph{{Infrared singularities of scattering
  amplitudes in perturbative QCD}},
  \href{https://doi.org/10.1103/PhysRevLett.102.162001,
  10.1103/PhysRevLett.111.199905}{\emph{Phys. Rev. Lett.} {\bfseries 102}
  (2009) 162001} [\href{https://arxiv.org/abs/0901.0722}{{\ttfamily
  0901.0722}}].

\bibitem{Gardi:2009qi}
E.~Gardi and L.~Magnea, \emph{{Factorization constraints for soft anomalous
  dimensions in QCD scattering amplitudes}},
  \href{https://doi.org/10.1088/1126-6708/2009/03/079}{\emph{JHEP} {\bfseries
  03} (2009) 079} [\href{https://arxiv.org/abs/0901.1091}{{\ttfamily
  0901.1091}}].

\bibitem{Almelid:2015jia}
O.~Almelid, C.~Duhr and E.~Gardi, \emph{{Three-loop corrections to the soft
  anomalous dimension in multileg scattering}},
  \href{https://doi.org/10.1103/PhysRevLett.117.172002}{\emph{Phys. Rev. Lett.}
  {\bfseries 117} (2016) 172002}
  [\href{https://arxiv.org/abs/1507.00047}{{\ttfamily 1507.00047}}].

\bibitem{Almelid:2017qju}
O.~Almelid, C.~Duhr, E.~Gardi, A.~McLeod and C.~D. White, \emph{{Bootstrapping
  the QCD soft anomalous dimension}},
  \href{https://doi.org/10.1007/JHEP09(2017)073}{\emph{JHEP} {\bfseries 09}
  (2017) 073} [\href{https://arxiv.org/abs/1706.10162}{{\ttfamily
  1706.10162}}].

\bibitem{Remiddi:1999ew}
E.~Remiddi and J.~A.~M. Vermaseren, \emph{{Harmonic polylogarithms}},
  \href{https://doi.org/10.1142/S0217751X00000367}{\emph{Int. J. Mod. Phys.}
  {\bfseries A15} (2000) 725}
  [\href{https://arxiv.org/abs/hep-ph/9905237}{{\ttfamily hep-ph/9905237}}].

\bibitem{Nogueira:1991ex}
P.~Nogueira, \emph{{Automatic Feynman graph generation}},
  \href{https://doi.org/10.1006/jcph.1993.1074}{\emph{J. Comput. Phys.}
  {\bfseries 105} (1993) 279}.

\bibitem{Chen:2019wyb}
L.~Chen, \emph{{A prescription for projectors to compute helicity amplitudes in
  D dimensions}},  \href{https://arxiv.org/abs/1904.00705}{{\ttfamily
  1904.00705}}.

\bibitem{Gehrmann:2013vga}
T.~Gehrmann, L.~Tancredi and E.~Weihs, \emph{{Two-loop QCD helicity amplitudes
  for $g\,g \to Z\,g$ and $g\,g \to Z\,\gamma $}},
  \href{https://doi.org/10.1007/JHEP04(2013)101}{\emph{JHEP} {\bfseries 04}
  (2013) 101} [\href{https://arxiv.org/abs/1302.2630}{{\ttfamily 1302.2630}}].

\bibitem{Peraro:2019cjj}
T.~Peraro and L.~Tancredi, \emph{{Physical projectors for multi-leg helicity
  amplitudes}}, \href{https://doi.org/10.1007/JHEP07(2019)114}{\emph{JHEP}
  {\bfseries 07} (2019) 114}
  [\href{https://arxiv.org/abs/1906.03298}{{\ttfamily 1906.03298}}].

\bibitem{Laporta:2001dd}
S.~Laporta, \emph{{High precision calculation of multiloop Feynman integrals by
  difference equations}}, \href{https://doi.org/10.1016/S0217-751X(00)00215-7,
  10.1142/S0217751X00002157}{\emph{Int. J. Mod. Phys.} {\bfseries A15} (2000)
  5087} [\href{https://arxiv.org/abs/hep-ph/0102033}{{\ttfamily
  hep-ph/0102033}}].

\bibitem{Vermaseren:2000nd}
J.~A.~M. Vermaseren, \emph{{New features of FORM}},
  \href{https://arxiv.org/abs/math-ph/0010025}{{\ttfamily math-ph/0010025}}.

\bibitem{Ruijl:2017dtg}
B.~Ruijl, T.~Ueda and J.~Vermaseren, \emph{{FORM version 4.2}},
  \href{https://arxiv.org/abs/1707.06453}{{\ttfamily 1707.06453}}.

\bibitem{Tkachov:1981wb}
F.~V. Tkachov, \emph{{A Theorem on Analytical Calculability of Four Loop
  Renormalization Group Functions}},
  \href{https://doi.org/10.1016/0370-2693(81)90288-4}{\emph{Phys. Lett.}
  {\bfseries 100B} (1981) 65}.

\bibitem{Chetyrkin:1981qh}
K.~G. Chetyrkin and F.~V. Tkachov, \emph{{Integration by Parts: The Algorithm
  to Calculate beta Functions in 4 Loops}},
  \href{https://doi.org/10.1016/0550-3213(81)90199-1}{\emph{Nucl. Phys.}
  {\bfseries B192} (1981) 159}.

\bibitem{Lee:2008tj}
R.~N. Lee, \emph{{Group structure of the integration-by-part identities and its
  application to the reduction of multiloop integrals}},
  \href{https://doi.org/10.1088/1126-6708/2008/07/031}{\emph{JHEP} {\bfseries
  07} (2008) 031} [\href{https://arxiv.org/abs/0804.3008}{{\ttfamily
  0804.3008}}].

\bibitem{Lee:2012cn}
R.~N. Lee, \emph{{Presenting LiteRed: a tool for the Loop InTEgrals
  REDuction}},  \href{https://arxiv.org/abs/1212.2685}{{\ttfamily 1212.2685}}.

\bibitem{Lee:2013hzt}
R.~N. Lee and A.~A. Pomeransky, \emph{{Critical points and number of master
  integrals}}, \href{https://doi.org/10.1007/JHEP11(2013)165}{\emph{JHEP}
  {\bfseries 11} (2013) 165} [\href{https://arxiv.org/abs/1308.6676}{{\ttfamily
  1308.6676}}].

\bibitem{Smirnov:2014hma}
A.~V. Smirnov, \emph{{FIRE5: a C++ implementation of Feynman Integral
  REduction}}, \href{https://doi.org/10.1016/j.cpc.2014.11.024}{\emph{Comput.
  Phys. Commun.} {\bfseries 189} (2015) 182}
  [\href{https://arxiv.org/abs/1408.2372}{{\ttfamily 1408.2372}}].

\bibitem{Kotikov:1991pm}
A.~V. Kotikov, \emph{{Differential equation method: The Calculation of N point
  Feynman diagrams}}, \href{https://doi.org/10.1016/0370-2693(91)90536-Y,
  10.1016/0370-2693(92)91582-T}{\emph{Phys. Lett.} {\bfseries B267} (1991)
  123}.

\bibitem{Bern:1993kr}
Z.~Bern, L.~J. Dixon and D.~A. Kosower, \emph{{Dimensionally regulated pentagon
  integrals}}, \href{https://doi.org/10.1016/0550-3213(94)90398-0}{\emph{Nucl.
  Phys.} {\bfseries B412} (1994) 751}
  [\href{https://arxiv.org/abs/hep-ph/9306240}{{\ttfamily hep-ph/9306240}}].

\bibitem{Gehrmann:1999as}
T.~Gehrmann and E.~Remiddi, \emph{{Differential equations for two loop four
  point functions}},
  \href{https://doi.org/10.1016/S0550-3213(00)00223-6}{\emph{Nucl. Phys.}
  {\bfseries B580} (2000) 485}
  [\href{https://arxiv.org/abs/hep-ph/9912329}{{\ttfamily hep-ph/9912329}}].

\bibitem{Henn:2013pwa}
J.~M. Henn, \emph{{Multiloop integrals in dimensional regularization made
  simple}}, \href{https://doi.org/10.1103/PhysRevLett.110.251601}{\emph{Phys.
  Rev. Lett.} {\bfseries 110} (2013) 251601}
  [\href{https://arxiv.org/abs/1304.1806}{{\ttfamily 1304.1806}}].

\bibitem{Henn:2016jdu}
J.~M. Henn and B.~Mistlberger, \emph{{Four-Gluon Scattering at Three Loops,
  Infrared Structure, and the Regge Limit}},
  \href{https://doi.org/10.1103/PhysRevLett.117.171601}{\emph{Phys. Rev. Lett.}
  {\bfseries 117} (2016) 171601}
  [\href{https://arxiv.org/abs/1608.00850}{{\ttfamily 1608.00850}}].

\bibitem{Henn:2019rgj}
J.~M. Henn and B.~Mistlberger, \emph{{Four-graviton scattering to three loops
  in $ \mathcal{N}=8 $ supergravity}},
  \href{https://doi.org/10.1007/JHEP05(2019)023}{\emph{JHEP} {\bfseries 05}
  (2019) 023} [\href{https://arxiv.org/abs/1902.07221}{{\ttfamily
  1902.07221}}].

\bibitem{Goncharov:1998kja}
A.~B. Goncharov, \emph{{Multiple polylogarithms, cyclotomy and modular
  complexes}}, \href{https://doi.org/10.4310/MRL.1998.v5.n4.a7}{\emph{Math.
  Res. Lett.} {\bfseries 5} (1998) 497}
  [\href{https://arxiv.org/abs/1105.2076}{{\ttfamily 1105.2076}}].

\bibitem{Panzer:2014gra}
E.~Panzer, \emph{{On hyperlogarithms and Feynman integrals with divergences and
  many scales}}, \href{https://doi.org/10.1007/JHEP03(2014)071}{\emph{JHEP}
  {\bfseries 03} (2014) 071} [\href{https://arxiv.org/abs/1401.4361}{{\ttfamily
  1401.4361}}].

\bibitem{Duhr:2011zq}
C.~Duhr, H.~Gangl and J.~R. Rhodes, \emph{{From polygons and symbols to
  polylogarithmic functions}},
  \href{https://doi.org/10.1007/JHEP10(2012)075}{\emph{JHEP} {\bfseries 10}
  (2012) 075} [\href{https://arxiv.org/abs/1110.0458}{{\ttfamily 1110.0458}}].

\bibitem{Panzer:2014caa}
E.~Panzer, \emph{{Algorithms for the symbolic integration of hyperlogarithms
  with applications to Feynman integrals}},
  \href{https://doi.org/10.1016/j.cpc.2014.10.019}{\emph{Comput. Phys. Commun.}
  {\bfseries 188} (2015) 148}
  [\href{https://arxiv.org/abs/1403.3385}{{\ttfamily 1403.3385}}].

\bibitem{Duhr:2012fh}
C.~Duhr, \emph{{Hopf algebras, coproducts and symbols: an application to Higgs
  boson amplitudes}},
  \href{https://doi.org/10.1007/JHEP08(2012)043}{\emph{JHEP} {\bfseries 08}
  (2012) 043} [\href{https://arxiv.org/abs/1203.0454}{{\ttfamily 1203.0454}}].

\bibitem{Maitre:2005uu}
D.~Maitre, \emph{{HPL, a mathematica implementation of the harmonic
  polylogarithms}},
  \href{https://doi.org/10.1016/j.cpc.2005.10.008}{\emph{Comput. Phys. Commun.}
  {\bfseries 174} (2006) 222}
  [\href{https://arxiv.org/abs/hep-ph/0507152}{{\ttfamily hep-ph/0507152}}].

\bibitem{Dixon:1996wi}
L.~J. Dixon, \emph{{Calculating scattering amplitudes efficiently}},  in
  \emph{{QCD and beyond. Proceedings, Theoretical Advanced Study Institute in
  Elementary Particle Physics, TASI-95, Boulder, USA, June 4-30, 1995}},
  pp.~539--584, 1996, \href{https://arxiv.org/abs/hep-ph/9601359}{{\ttfamily
  hep-ph/9601359}},
  \href{http://www-public.slac.stanford.edu/sciDoc/docMeta.aspx?slacPubNumber=SLAC-PUB-7106}{http://www-public.slac.stanford.edu/sciDoc/docMeta.aspx?slacPubNumber=SLAC-PUB-7106}.

\bibitem{Herzog:2017}
F.~Herzog, B.~Ruijl, T.~Ueda, J.~A.~M. Vermaseren and A.~Vogt, \emph{{The
  five-loop beta function of Yang-Mills theory with fermions}},
  \href{https://doi.org/10.1007/JHEP02(2017)090}{\emph{JHEP} {\bfseries 02}
  (2017) 090} [\href{https://arxiv.org/abs/1701.01404}{{\ttfamily
  1701.01404}}].

\bibitem{Baikov:2016}
P.~A. Baikov, K.~G. Chetyrkin and J.~H. Kühn, \emph{{Five-Loop Running of the
  QCD coupling constant}},
  \href{https://doi.org/10.1103/PhysRevLett.118.082002}{\emph{Phys. Rev. Lett.}
  {\bfseries 118} (2017) 082002}
  [\href{https://arxiv.org/abs/1606.08659}{{\ttfamily 1606.08659}}].

\bibitem{Bern:1990ux}
Z.~Bern and D.~A. Kosower, \emph{{Color decomposition of one loop amplitudes in
  gauge theories}},
  \href{https://doi.org/10.1016/0550-3213(91)90567-H}{\emph{Nucl. Phys.}
  {\bfseries B362} (1991) 389}.

\bibitem{Naculich:2011ep}
S.~G. Naculich, \emph{{All-loop group-theory constraints for color-ordered
  SU(N) gauge-theory amplitudes}},
  \href{https://doi.org/10.1016/j.physletb.2011.12.010}{\emph{Phys. Lett.}
  {\bfseries B707} (2012) 191}
  [\href{https://arxiv.org/abs/1110.1859}{{\ttfamily 1110.1859}}].

\end{thebibliography}\endgroup

\end{document}